\documentclass[runningheads]{llncs}
\usepackage[T1]{fontenc}
\usepackage{graphicx}
\usepackage{hyperref}
\usepackage{color}

\usepackage{pifont}
\usepackage{tabularx}
\usepackage{algorithm}
\usepackage{algpseudocode}
\usepackage{placeins}
\usepackage{subcaption}
\usepackage{comment}

\usepackage{alphabeta}  

\algrenewcommand\algorithmicthen{}	

\definecolor{darkgreen}{rgb}{0.1, 0.6, 0.1}

\usepackage{ifthen}
\usepackage{calc}
\usepackage{color}
\usepackage{tikz}
\usepackage{amsfonts}
\usepackage{pgfplots}

\DeclareRobustCommand{\change}[4]{{\color{#3}#1}\ifx\relax#4\relax\else\xspace\textcolor{Gray}{\sout{#4}}\fi}

\begin{document}
\title{Adaptive Overlay for Fast Peer-to-Peer Dissemination}
%
%
\author{Evangelos Kolyvas\orcidID{0000-0003-1939-1120}\\
Spyros Voulgaris\orcidID{0000-0003-3023-0299}}
%
\authorrunning{E. Kolyvas and S. Voulgaris}
%
\institute{Department of Informatics,\\
Athens University of Economics and Business,\\
Athens, Greece\\
\email{\{ekolyvas,voulgaris\}@aueb.gr}}
%
\maketitle              
%
\begin{abstract}

Efficient message dissemination is a fundamental requirement in many
	Peer-to-Peer (P2P) systems, where a message generated by one node must be
	propagated rapidly and reliably throughout the network.
The time required for network-wide dissemination is strongly influenced by the
	topology of the underlying P2P overlay and, in particular, by the quality of
	the links through which messages are forwarded.
In many systems, overlay connections are established through random peer
	selection, which provides robustness and scalability but does not explicitly
	account for the observed performance or network proximity of neighboring
	nodes.

This paper presents an adaptive decentralized overlay construction approach for
	accelerating message dissemination in P2P networks.
The approach dynamically selects and maintains neighbors using three
	complementary principles:
(i) a performance-based selection mechanism that evaluates how effectively
	neighboring peers contribute to message dissemination,
(ii) a latency-based selection mechanism that favors low-latency connections,
	and
(iii) random peer selection that maintains global connectivity and enables the
	discovery of previously unseen peers.
By combining these mechanisms, the resulting overlay simultaneously exploits
	high-performing links, low-latency connections, and long-range shortcuts.
The approach is applicable to a broad range of P2P systems in which nodes
	disseminate messages throughout the network, including distributed ledgers
	as one representative example.

\end{abstract}

\section{Peer-to-Peer Dissemination Background}
\label{sec:background}

Peer-to-Peer (P2P) systems consist of distributed nodes that communicate
directly with one another without relying on a centralized communication
infrastructure.
A common requirement in such systems is the dissemination of information
generated by one node to a large fraction, or the entirety, of the
network~\cite{cliquesensus,scramble,cougar}.
Many P2P dissemination networks rely on unstructured overlay networks, in which
communication is typically based on gossiping~\cite{epidemic1,epidemic2}.
Since P2P networks can be highly dynamic and may consist of a large number of
nodes, it is infeasible for each node to maintain complete knowledge of the
network membership.
Instead, each node maintains links to a relatively small number of other nodes
in the network.
Two nodes are referred to as \emph{neighbors} if they are connected by a link.

In many P2P dissemination networks, the \emph{peer selection mechanism}, used to
determine a node's neighbors, relies on randomly selecting
peers~\cite{randomized} from a set of candidate nodes provided by lower-level
protocols that offer peer discovery services~\cite{peerSampling}.
This random peer selection approach provides a simple and robust way of
maintaining connectivity without requiring nodes to maintain global knowledge of
the network.
However, random peer selection does not explicitly account for the observed
performance or network proximity of neighboring nodes~\cite{peerSelection}.
Consequently, nodes that consistently deliver messages slowly may remain in a
node's neighborhood, while nodes that can disseminate messages more efficiently
may remain underutilized.
Furthermore, relying exclusively on random connections may result in
unnecessarily long paths between nodes that are in close proximity.

These observations motivate an adaptive approach in which the overlay topology
is continuously adjusted according to both the observed contribution of
neighboring nodes to message dissemination and the latency of the underlying
network.
Such an approach can combine high-performing connections, low-latency paths, and
random long-range connections to reduce network-wide message propagation delay
while preserving connectivity and enabling the discovery of potentially better
peers.

\section{Adaptive Overlay Construction}
\label{sec:mechanisms_intro}

We introduce an adaptive decentralized overlay construction approach that lets
each node dynamically determine which peers to maintain as neighbors.

Each node maintains a fixed-size neighborhood that is divided into two disjoint
sets: a \emph{performance set} and a \emph{latency set}.
The performance set contains peers selected according to their observed
contribution to message dissemination, while the latency set contains peers
selected according to their measured network latency.
In addition, each set includes a number of randomly selected peers that are
periodically replaced to enable continuous exploration of the network.

The two selection mechanisms optimize complementary properties of the overlay.
The performance-based selection identifies peers that are consistently effective
at delivering newly disseminated messages, while the latency-based selection
identifies peers that can be reached through low-latency network paths.
The randomly selected peers provide continuous exploration of the network,
allowing each node to discover previously unseen peers that may offer better
dissemination performance or lower latency.

The following sections describe the two adaptive peer-selection mechanisms.

\subsection{Performance-Based Selection}
\label{sec:performance_based}

A node builds its performance set by applying the performance-based selection
mechanism.
The objective is to identify neighbors that consistently contribute to fast
message dissemination.

For each newly disseminated message, the node observes the order in which the
message arrives from different peers in its performance set.
Suppose $t_1$ is the time at which the node receives the message for the first
time from peer $u$, where $u$ belongs to the performance set, and $t_2$ is the
time at which the same message is subsequently received from another peer $v$,
where $v$ also belongs to the performance set and $u \neq v$.
The node adds the difference between these timestamps, measured in milliseconds,
to the score of the first peer $u$:

\[
score(u) \leftarrow score(u) + (t_2 - t_1)
\]

Both $t_1$ and $t_2$ refer to the local clock of the receiving node.
A large difference indicates that the message arrived significantly earlier
through $u$ than through another neighbor, suggesting that $u$ provides a useful
dissemination path.
The peer that delivers the message first therefore receives credit, while the
other peers in the performance set receive no points for that message.

Periodically, every $k$ messages, the node ranks the peers in its performance
set according to their cumulative scores over the previous $k$ messages.
It retains the highest-scoring peers as stable members of the performance set,
removes the lowest-scoring peers, and replaces them with randomly selected peers
from the underlying peer discovery service.
This random replacement mechanism continuously introduces previously unseen
peers into the performance set, allowing the node to discover peers that may
perform better than its current neighbors.

The scores of all peers in the updated performance set are then reset to zero,
including peers retained from the previous evaluation period.
The process is repeated every $k$ messages.

The main advantage of the performance-based selection mechanism is that its
objective is directly derived from the dissemination process itself.
Rather than relying on manually selected characteristics of individual nodes or
on a static topology, the selection mechanism evaluates peers according to their
observed contribution to message delivery.

As a result, the performance-based selection mechanism can implicitly capture
multiple factors that influence dissemination performance, including network
latency, available bandwidth, congestion, routing conditions, node processing
capacity, and other application- or system-specific factors that affect how
quickly a peer can receive and relay messages.
Importantly, these factors do not need to be explicitly modeled.
Their combined effect is reflected in the observed message arrival times.

The performance-based selection mechanism therefore adapts to heterogeneous
network conditions without requiring prior knowledge of the characteristics of
the underlying P2P environment.

\subsection{Latency-Based Selection}
\label{sec:latency_based}

The performance-based selection mechanism evaluates peers according to their
observed contribution to message dissemination.
A complementary mechanism is required to explicitly exploit low-latency network
connections.

For the latency set, the approach applies a latency-based selection mechanism
that operates at the networking/P2P layer.
Unlike the performance-based selection mechanism, it does not depend on the
dissemination behavior of the application.
Instead, it measures the network latency between the node and each peer in its
latency set.

The node periodically sends ping messages to the peers in its latency set and
measures their corresponding round-trip times (RTTs).
After every measurement period, it calculates the average latency for each peer,
retains the peers with the lowest average latency, and removes the peers with
the highest average latency.
The removed peers are replaced with randomly selected peers from the underlying
peer discovery service.

As with the performance-based selection mechanism, random replacement is
important because it continuously exposes the node to previously unseen peers.
A peer that is initially unknown to the node may provide a substantially
lower-latency connection than the current neighbors and can therefore be
discovered through this exploration process.

All current latency measurements are discarded after each evaluation period,
including those corresponding to peers that remain in the latency set.
The measurement process then starts again.
This periodic rejuvenation enables the overlay to adapt to changing network
conditions, such as variations in congestion, routing paths, or connectivity.

Although low-latency connections often correspond to geographically nearby
nodes, the latency-based selection mechanism does not explicitly use
geographical location.
It relies solely on measured network latency and therefore remains applicable
across heterogeneous network environments.
Moreover, the latency set is formed asynchronously with respect to message
dissemination, allowing low-latency links to be discovered even during periods
of sporadic message dissemination.

\subsection{Combining Local and Global Dissemination}
\label{sec:local_global}

In general, rapid dissemination at a global scale has to address two
complementary requirements.

First, a newly generated message should be distributed quickly within a local
region of the network.
Low-latency connections can facilitate this process by allowing the message to
move rapidly among nodes that can communicate efficiently with one another.
Such connections can reduce the delay required to exhaustively cover nearby
portions of the network.

Second, a dissemination topology should also provide efficient paths between
distant regions of the network.
If dissemination relies exclusively on local connections, a message may have to
traverse a large number of intermediate nodes before reaching remote parts of
the network.
Long-range connections can provide shortcuts that significantly reduce this
delay.
The randomly selected peers in both sets provide such long-range connections,
while also enabling continuous exploration of the network.

For example, consider a message generated by a node in New York.
A low-latency connection can allow the message to reach a nearby node in Boston
quickly.
At the same time, a long-range connection from a node in the New York region to
a node in Frankfurt can allow the message to reach that remote region without
waiting for it to propagate gradually through numerous intermediate nodes across
North America and Europe.\newline

In summary, the combination of performance-based selection, latency-based
selection, and random peer selection allows the overlay to exploit
high-performing links, low-latency paths, and long-range connections.
This combination enables the topology to adapt to both observed dissemination
performance and changing network conditions.

\section{Conclusions}
\label{sec:conclusions}

We presented an adaptive decentralized overlay construction approach for
accelerating network-wide message dissemination in Peer-to-Peer networks.
The approach dynamically determines which neighbors each node should maintain by
combining three complementary principles:
\emph{(a) the observed recent performance of neighboring peers in delivering new
messages},
\emph{(b) network proximity} in terms of measured latency, and
\emph{(c) random} peer selection.

The performance-based selection mechanism favors links that consistently
contribute to early message delivery, while the latency-based mechanism
explicitly exploits low-latency communication paths.
Random peer selection complements these mechanisms by maintaining network
diversity, introducing long-range shortcuts, and enabling the continuous
discovery of previously unseen peers.

The proposed approach is general and does not depend on a particular
application-layer protocol.
It can therefore be applied to P2P systems in which nodes are required to
disseminate messages throughout the network.
Distributed ledger networks constitute one representative application, but the
same principles can be applied to other P2P dissemination workloads.

Finally, the adaptive nature of the overlay allows it to respond to
heterogeneous and changing network conditions without requiring the topology to
be manually optimized for a particular deployment.

\section*{Acknowledgment}

Work funded by \emph{Input Output Global} (\emph{IOG}) in the context of the
\emph{``Eclipse-Resistant Network Overlays for Fast Data Dissemination''}
project, to optimize Cardano's overlay network.

\newpage
\bibliographystyle{splncs04}
\bibliography{refs}

\end{document}